# EXTENDING DEEP LEARNING U-NET ARCHITECTURE FOR PREDICTING UNSTEADY FLUID FLOWS IN TEXTURED MICROCHANNELS

**Ganesh Sahadeo Meshram[1], Partha P. Chakrabarti[2], Suman Chakraborty[1*]**

[1]Department of Mechanical Engineering, IIT Kharagpur, Kharagpur, 721302, India
[2]Department of Computer Science & Engineering, IIT Kharagpur, Kharagpur, 721302, India

## ABSTRACT

In this study, we have explored an application of deep learning architecture of the U-Net model, originally designed for biomedical image segmentation, in a regression analysis aimed at predicting fluid flows through textured microchannels. The data for this analysis is generated using the lattice Boltzmann method through extensive simulations, capturing the intricate behaviors of fluid dynamics in a microscale environment. The raw simulation data was meticulously preprocessed to prepare it for training the U-Net model, ensuring that the input features and labels were appropriately formatted and normalized to optimize the learning process of the model. The U-Net model, with its inherent capability of capturing spatial hierarchies and producing better predictions, proved effective in this novel application. We have evaluated the performance of the model using metrics including MSE, RMSE, MAE, and $R^2$ scores. These metrics were crucial in assessing the accuracy and reliability of the model predictions. The results demonstrate that the U-Net model can predict fluid flows with high accuracy and less error, indicating its potential for broader applications in fluid dynamics and other fields requiring precise regression modeling. A parametric analysis of the U-Net with attention mechanism showed that the velocity field prediction is contingent upon the solid-fluid interaction parameter and surface wettability. The U-Net equipped with an attention mechanism predicts the velocity magnitude and components for textured microchannels with an average error of 5.18%, which upon optimization may subsequently lower to 2.1%. The U-Net model including an attention mechanism (U-Net AM) regularly surpasses the conventional U-Net model in all measures, evidencing enhanced accuracy and generalization. This study highlights the versatility of U-Net beyond its conventional use in image segmentation, showing its applicability in complex regression tasks. The developed model can be a valuable tool in predicting fluid behavior in various domains where accurate modeling of fluid flow is essential.

## 1. INTRODUCTION

Fluid flow in textured microchannels exhibits complex behavior due to surface patterns like grooves or protrusions. Recent studies have revealed that texturing can reduce flow resistance by creating apparent slip conditions [1]. Introducing surface textures leads to intricate flow structures, including vortices and boundary layer variations. A recent study has also explored the effects of convergence-divergence angles in hourglass-shaped microchannels on pressure drop and flow physics [2]. The fluid dynamics in microchannels are significantly different from those in macroscale channels due to the dominance of viscous forces, surface tension, and other microscale phenomena [3–7]. When fluid flows with time-varying velocity and pressure fields occur in textured microchannels, the prediction of these flows becomes increasingly complex and computationally expensive [8]. Accurate prediction of such fluid flows is essential for optimizing microfluidic device performance, making this an important area of research [9]. Conventional computational fluid dynamics (CFD) models have been widely used to simulate fluid flows in microchannels [10]. However, these models often require significant computational resources, especially for flows in channels with complex geometries such as textured surfaces [11]. The need for high-resolution simulations over long time period further increases the computational cost. To address these limitations, researchers have turned to machine learning techniques, particularly deep learning, which can provide faster and potentially more accurate predictions without the heavy computational costs associated with CFD [12, 13].

---

*Corresponding Author: suman@mech.iitkgp.ac.in

In recent years, deep learning models have shown great promise in addressing complex fluid dynamics problems. Convolutional Neural Networks (CNNs) have been particularly successful in this regard due to their ability to capture spatial hierarchies in data [14]. Among the CNN-based architectures, the deep learning model has gained attention for its utility in fluid flow prediction [15]. The U-Net model has a unique encoder-decoder structure that allows it to capture both global and local features in the input data, making it highly effective for tasks involving spatially complex and structured data, such as fluid flow prediction. U-Net has been applied to various fluid dynamics problems, demonstrating impressive results in predicting steady-state and laminar flows. For example, Luo et al. [16] applied a modified CNN architecture to predict steady-state fluid flows in two-dimensional geometries, showing that the model could accurately predict flow velocity and pressure fields with significantly reduced computational time compared to CFD simulations. This success in steady-state problems has led to increased interest in extending the capabilities of the U-Net model to predict flows, particularly in microfluidic systems where real-time or near-real-time predictions are often required. Textured microchannels add another layer of complexity to the problem of flow prediction [17]. The surface textures in these channels can induce additional flow phenomena such as slip, recirculation, and turbulence, which significantly alter the overall flow behavior. These interactions are difficult to model using conventional methods due to the multiscale nature of the problem, where small-scale surface textures influence the larger-scale flow dynamics. Incorporating these surface interactions into predictive models is essential for the accurate simulation of fluid flows in textured microchannels. Extending U-Net for the prediction of fluid flows in textured microchannels involves several key modifications to the architecture. First, temporal dynamics must be incorporated into the model, as flows are inherently time-dependent. This can be achieved by integrating recurrent neural networks (RNNs) or long short-term memory (LSTM) networks into the U-Net architecture to capture the time-evolving nature of the flow fields [15]. Second, surface texture information must be explicitly included in the input data. This can be done by augmenting the input with geometric features or by using multiple input channels to provide the model with detailed information about the surface of the microchannel. The growing body of research on deep learning for fluid dynamics suggests that such approaches can lead to more efficient and accurate predictions compared to traditional CFD methods. For instance, Kim et al. [18] demonstrated the efficacy of deep learning models in predicting turbulent flows in porous media, a problem that shares many similarities with flow in textured microchannels. The success of these models in capturing complex flow behaviors in other contexts makes them a promising tool for tackling the challenges of predicting fluid flows in textured microchannels.

Recent advancements in deep learning have shown significant potential in addressing complex fluid dynamics problems, particularly in predicting fluid flows. Convolutional neural networks (CNNs), especially the U-Net architecture, have become a preferred choice for fluid flow prediction due to their ability to handle spatial complexity. Guo et al. [19] applied hybrid deep learning methods to analyze and predict steady-state flow predictions in two- dimensional geometries, showing reduced computational time compared to traditional CFD methods. Lu et al. [20] extended CNNs for laminar and turbulent flow predictions, emphasizing the role of deep learning in modeling flows. Several researchers have focused on integrating recurrent neural networks (RNNs) and long short-term memory (LSTM) networks with CNNs to predict time-dependent flows. Lee et al. [21] incorporated LSTM into U-Net for capturing temporal dynamics in flows, while Xie et al. [22] enhanced CNN performance by adding RNN layers for time-series predictions in complex geometries. In the context of textured microchannels, Luo et al. [20] investigated the impact of surface textures on flow behavior, suggesting that deep learning models could efficiently predict complex interactions between flow and surface textures. Moreover, the development of physics-informed neural networks (PINNs) has enabled models to respect physical laws while predicting fluid flows [23]. Raissi et al. [24] introduced PINNs for solving Navier-Stokes equations, while Jin et al. [25] applied them to microfluidic flows. Other notable works include the efforts of Chang et al. [26], who used transfer learning for fluid flow predictions, and Montans et al. [27], who developed data-augmentation techniques to enhance model generalization in microchannels. Each of these contribution highlights the transformative potential of deep learning in fluid dynamics. In this work, we employ U-Net deep learning architecture, to perform regression analysis for predicting fluid flow variables in textured microchannels. The study utilizes data produced by the lattice Boltzmann method, demonstrating complex fluid dynamics at the microscale. The data was carefully preprocessed to enhance the format and standardize the features and labels for the U-Net model, hence augmenting its training efficiency.

## 2. PROBLEM DEFINITION AND DATA COLLECTION

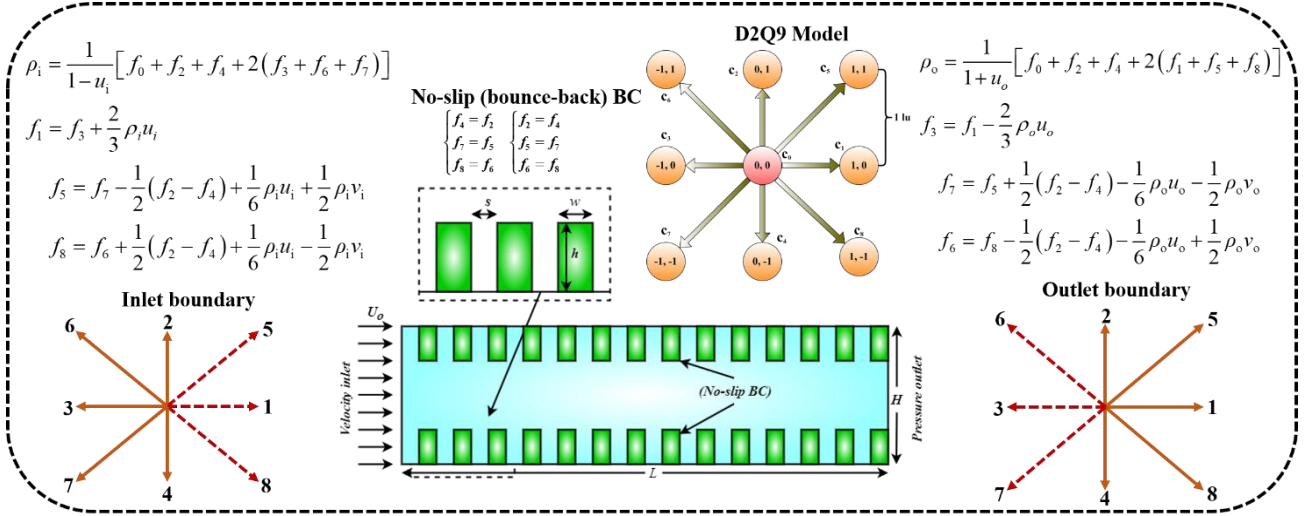

**Fig. 1** A 2-D fluid flow through textured microchannels. The textures, characterized by height h, width w, and inter-texture spacing s, are positioned at the upper and lower walls. The dimensions of the microchannel are designated as L for length and H for height. Uniform velocity boundary conditions are implemented at the channel entrance, while pressure boundary conditions are imposed at the exit.

Figure 1 shows the problem for analyzing 2-D fluid flow through a microchannel with textured surfaces. The textures with height h, width w, and spacing between two textures s, are imposed at the top and bottom walls. The length and height of microchannel are taken as L, and H respectively. In this case, we have considered the surfaces to be hydrophobic with no-slip boundary conditions. Uniform velocity inlet and pressure outlet BCs are applied at the channel entrance and exit, respectively. Simulations of this fluid flow through textured microchannel problem are done using the pseudo-potential multiphase lattice Boltzmann method with the D2Q9 model and the dataset is collected to provide dataset for training of U-Net models.

### 2.1 Governing equations
In the case of an incompressible Newtonian fluid, the governing equations reflect the conservation of mass and the conservation of momentum, respectively:

$$\frac{\partial \rho}{\partial t} + \nabla \cdot (\rho \mathbf{u}) = 0 \tag{1}$$

$$\rho \left( \frac{\partial \mathbf{u}}{\partial t} + (\mathbf{u} \cdot \nabla) \mathbf{u} \right) = -\nabla p + \mu \nabla^2 \mathbf{u} + \mathbf{F} \tag{2}$$

where **ρ** is the density of the fluid, *p* is the pressure, **u** is the velocity vector, and **F** is the body force. For a Newtonian fluid, the stress tensors can be expressed as $\Delta \cdot \tau = \mu \nabla^2 \mathbf{u}$ with μ being the fluid viscosity.

### 2.2 Lattice Boltzmann method
The lattice Boltzmann equation with a single-relaxation-time collision operator is formulated as [28]

$$f_p(x + e_p \delta t) = f_p(x,t) + \frac{\delta t}{\tau} \left[ f_p(x,t) - f_p^{eq}(x,t) \right] \tag{3}$$

$$f_p^{eq}(x,t) = \rho \omega_p \left[ 1 + \frac{e_p \cdot u}{c_{ls}^2} + \frac{(e_p \cdot u)^2}{2c_{ls}^2} - \frac{u^2}{2c_{ls}^2} \right] \tag{4}$$

Here, $f_p$ represents the density distribution function, $f_p^{eq}$ denotes the equilibrium density function, *x* signifies the spatial coordinate, $e_p$ indicates the particle velocity in direction *p*, *p* = 0,1,2,3,....,n, $\omega_p$ is the weighting coefficient, $c_{ls} = \frac{c}{\sqrt{3}}$ refers to the sound speed of a lattice, and $c = \frac{\Delta x}{\Delta t}$ represents the lattice constant, where $\Delta x$ and $\Delta t$ are lattice spatial step and lattice timesteps, respectively, **u** is the macroscopic velocity (vector form) [28]. The lattice Boltzmann equation is resolved through two stages, i.e., collision and streaming.

$$f_p^{out}(x,t) = f_p\left(x + e_p \delta t\right) = f_p^{in}(x,t) - \frac{1}{\tau}\left[f_p^{in}(x,t) - f_p^{eq}(x,t)\right] \tag{5}$$

$$f_p^{in}\left(x + e_p \delta t, t + \delta t\right) = f_p^{out}(x, t + \delta t) \tag{6}$$

The macroscopic density and velocity can be determined [29] using the distribution function as follows:

$$\rho = \sum_p f_p = \sum_p f_p^{eq} \tag{7}$$

$$u = \frac{1}{\rho}\sum_p f_p \cdot e_p = \sum_p f_p^{eq} \cdot e_p \tag{8}$$

The change in velocity and the actual fluid velocity due to total force are calculated as

$$\Delta u = \frac{F \delta t}{2\rho} \tag{9}$$

$$U = u + \frac{F \delta t}{2\rho} \tag{10}$$

### 3. DEEP LEARNING APPROACH USING U-NET MODEL

**3.1 U-Net Model**

Figure 2 depicts a U-Net architecture modified for regression tasks, including the prediction of unstable fluid dynamics. The architecture comprises an encoder-decoder framework including skip connections. The encoder (left side) systematically lowers spatial dimensions while increasing feature depth. The process begins with a 664 input, denoting four variables (e.g., spatial coordinates x, y, pressure p, density ρ), and employs convolutional layers that double the filter count at each stage (64, 128, 256, 512), concurrently downsampling the spatial resolution. The deepest feature map, 256512, is sent to the decoder (right side), where transposed convolutions reinstate the spatial dimensions while reducing the number of filters by half at each stage. Skip connections merge the feature maps of encoders with the matching decoder layers, enabling the model to retrieve intricate information. The output layer generates a 12801 feature map, predicting two continuous variables, and presumably velocity components, making this model appropriate for regression-based prediction in fluid flows.

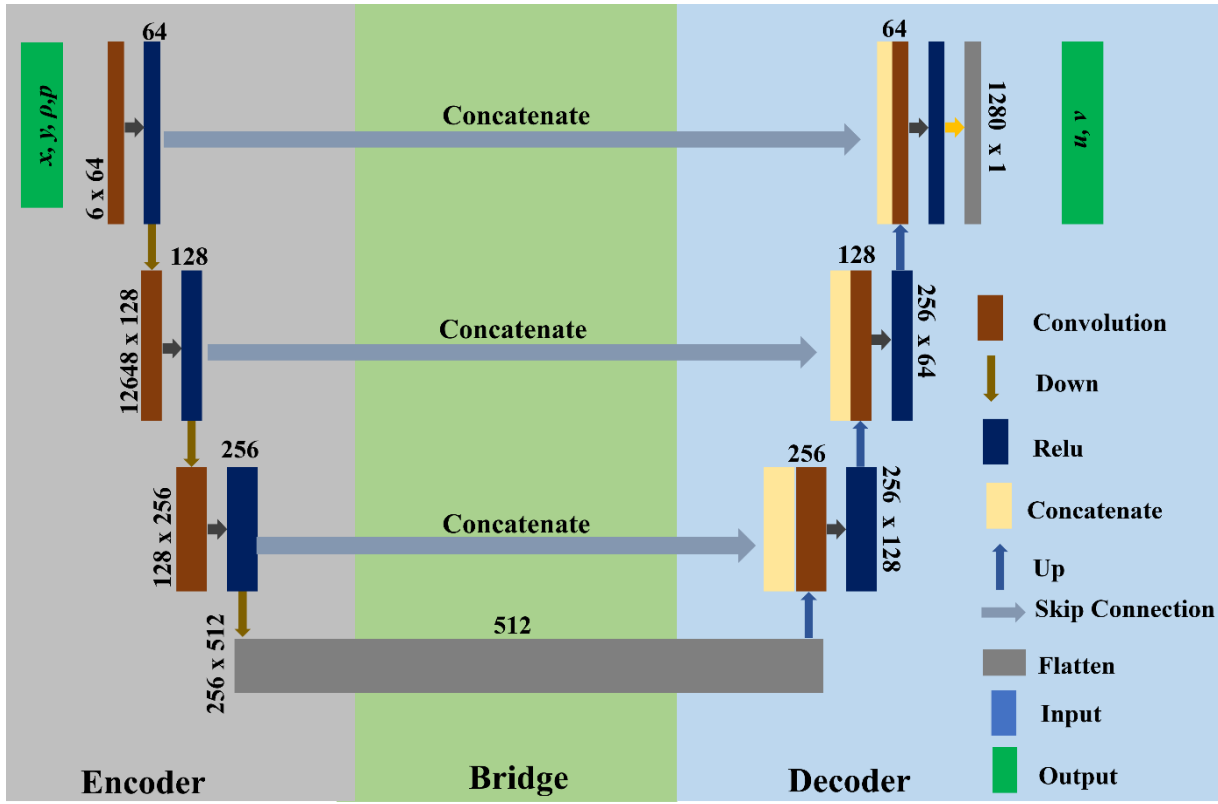

**Fig. 2** Prediction of unsteady velocity profiles using U-Net model for different timesteps a) Actual prediction

### 3.2 Methodology
The comprehensive methodology illustrates a complete fluid flow analysis framework through multiple interconnected stages as shown in Fig. 3. Figure 3(a) presents the problem definition with a geometrical setup of periodic obstacles in a channel flow configuration. Figure 3(b) demonstrates the CFD approach utilizing the D2Q9 Lattice Boltzmann model with nine discrete velocity directions, while Fig. 3(c) shows the governing equations including continuity and Navier-Stokes equations with their corresponding density and velocity formulations. The simulation results in Fig. 3(d) display the true velocity fields (X and Y components) and vorticity distribution, which serve as the reference data. Figure 3(e) showcases the segmented data input spanning different X-ranges (0 − 200, 200 − 400, 400 − 600, 600 − 800) with corresponding velocity fields, leading to Fig. 3(f) data preprocessing stage that ensures proper integration of the computational results. Figure 3(g) illustrates the deep learning architectures, comparing standard U-Net and enhanced U-Net AM (with attention mechanism) implementations, while Fig. 3(h) presents the prediction results for both velocity components and vorticity fields. Finally, Fig. 3(i) provides error estimation analysis, including spatial error distribution maps and convergence plots for both neural network models, demonstrating the quantitative accuracy of the predictions through L2 norm calculations and epoch-wise error.

### 3.3 Model training
Table 1 delineates the principal hyperparameters used in a U-Net Attention Model, offering a comprehensive understanding of the model architecture and training configuration. The input shape is specified as (None, 10, 6), with None indicating a configurable batch size, and the model handling 10 phases, each including 6 features. The training is performed with a batch size of 32, using the Adam optimizer with a learning rate of 0.001. ReLU and

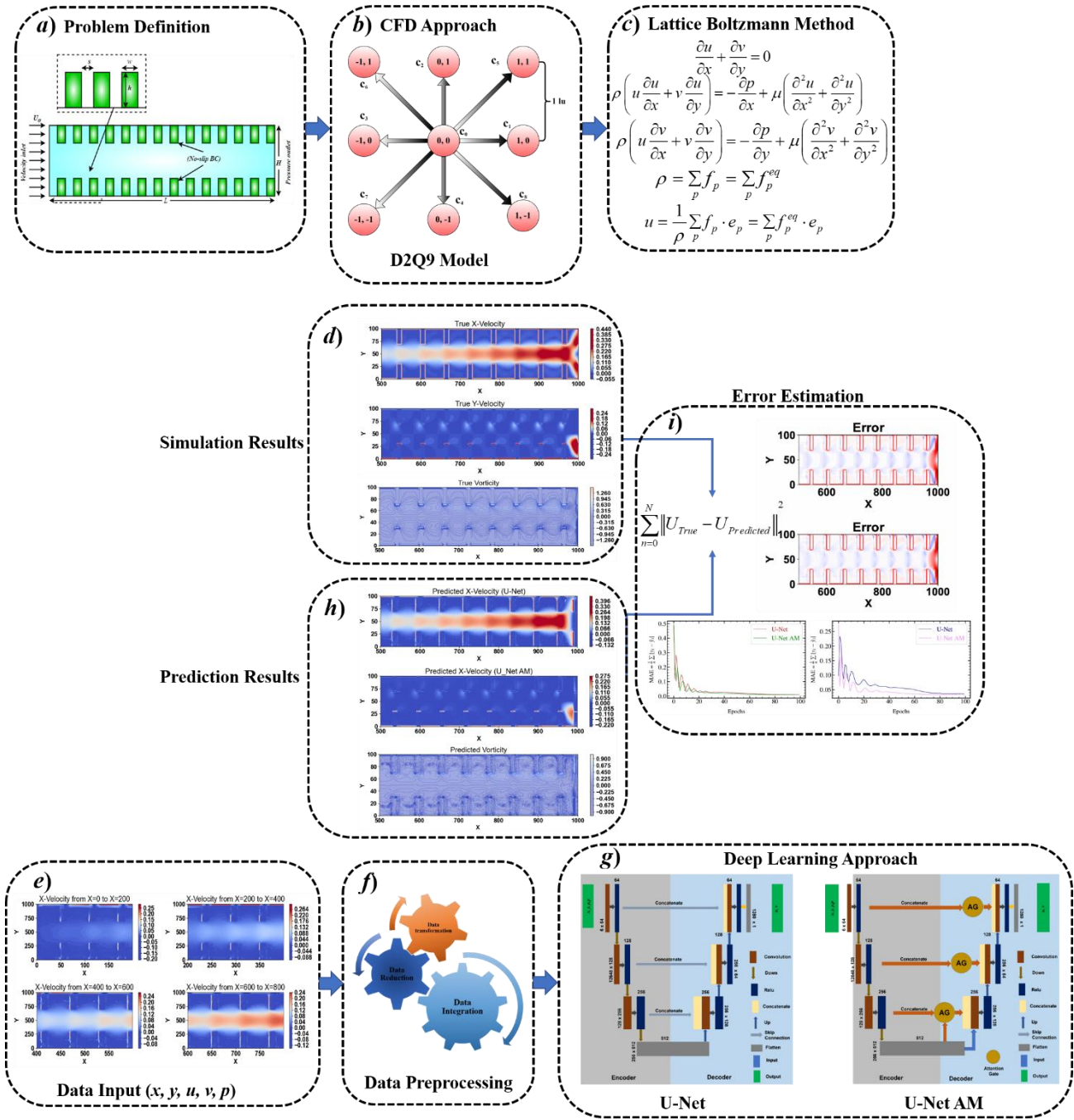

**Fig. 3** Proposed model architecture of the attention U-Net.

Softmax serve as activation functions, whilst binary cross-entropy is used as the loss function for optimization. The model is trained for 50 epochs, with a dropout rate of 0.3 for regularization. The attention mechanism employs additive attention to improve feature selection. The design has 17 dense layers, with diverse units (64, 128, and 256), hence providing enough capacity for intricate learning tasks. This setup optimizes speed, flexibility, and regularization for efficient training of U-Net models.

**Table 1** Hyperparameters used in U-Net Model.

| Hyperparameter | Value |
| --- | --- |
| Input shape | (None, 10, 6) |
| Batch size | 32 |
| Optimizer | Adam |
| Learning rate | 0.001 |
| Activation function | ReLU, Softmax |
| Loss function | Binary cross- entropy |
| Number of epochs | 50 |
| Dropout rate | 0.3 |
| Attention mechanism | Additive attention |
| Number of dense layer | 17 |
| Units in dense layer | (64, 128, 256) |

## 4. RESULTS AND DISCUSSION

**4.1 Performance evaluation**

The performance of models and the accuracy of the prediction has assessed using MAE, MSE, RMSE, and R² scores, which can be defined using the following formulas:

$$MAE = \frac{1}{n}\sum_{i=1}^{n}|y_i - \hat{y}_i| \quad (11)$$

$$MSE = \frac{1}{n}\sum_{i=1}^{n}(y_i - \hat{y}_i)^2 \quad (12)$$

$$RMSE = \sqrt{\frac{1}{n}\sum_{i=1}^{n}(y_i - \hat{y}_i)^2} \quad (13)$$

$$R^2 = 1 - \frac{\sum_{i=1}^{n}(y_i - \hat{y}_i)^2}{\sum_{i=1}^{n}(y_i - \overline{y})^2} \quad (14)$$

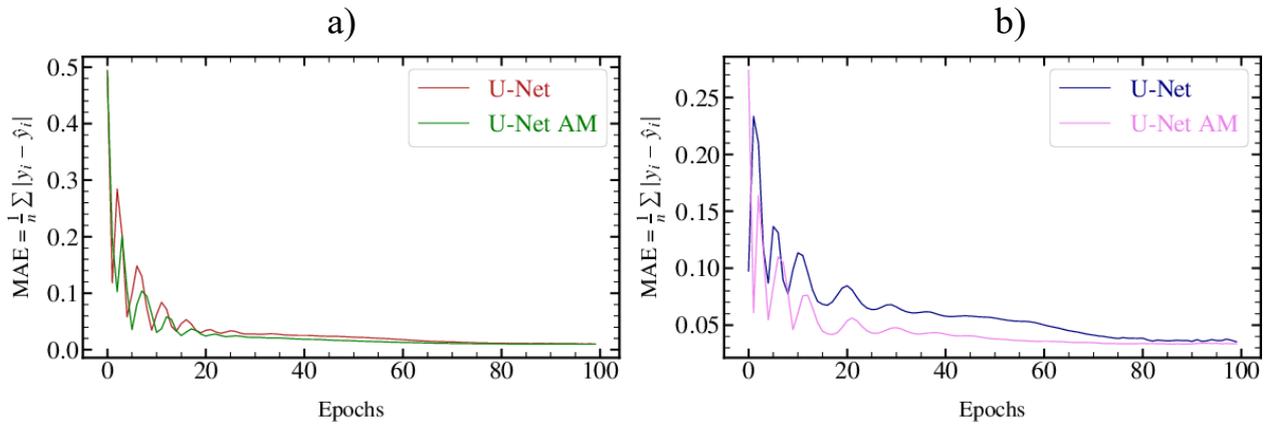

**Fig. 4** Loss for training and validation of fluid flow datasets. a) Training, and b) Validation

Figure 4 shows loss using mean squared error to illustrate the training and validation loss for both u and v predictions throughout 100 epochs. In the first plot, the training loss begins at roughly 0.45 and swiftly declines to around 0.02 by epoch 40, thereafter exhibiting stability. The validation loss commences at 0.42, converging to around 0.03 by epoch 40, with minor variations afterward. This signifies effective convergence for both training and validation, illustrating the capacity of model to generalize proficiently for velocity predictions. In the second plot (loss for u), the training loss commences at around 0.6 and declines more steeply, reaching roughly 0.03 by epoch 40. The validation loss exhibits a similar pattern, starting at around 0.22 and decreasing to almost 0.025 by epoch 40. Subsequently, both losses stabilize with little variations. The prevailing patterns indicate that the model is proficiently reducing loss for both u and v predictions, demonstrating robust generalization performance, since there is no indication of overfitting across the training and validation sets after 100 epochs.

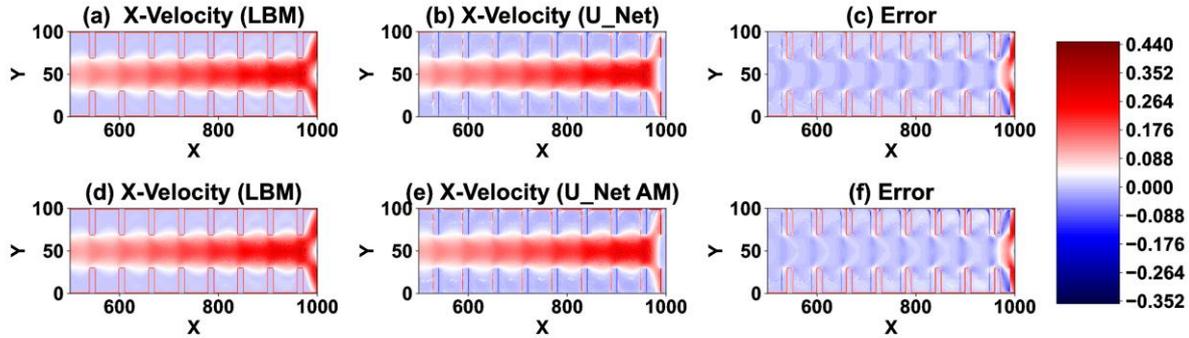

**Fig. 5** Prediction of x-velocity component using U-Net and attention U-Net (testing dataset).

Streamwise and transverse velocity fields predicted by U-Net and U-Net with an attention mechanism by flow fields with LBM simulations, highlighting their accuracy through mean absolute error as shown in Fig. 5. Figure 5-(a,b,c) shows the prediction of streamwise velocity field taken as a reference from the LBM simulation, and U-Net captures broad flow features and exhibits significant errors near textures with an error of ±0.176. The attention mechanism-based U-Net model improves accuracy, reducing errors to ±0.088 shown in Figure 5-(d,e,f). Figure 6 displays y-velocity fields, with the LBM solution showing complex flow patterns around textures. U-Net diminishes near boundaries, resulting in higher errors (ref to Fig. 6-(a,b,c)), while attention U-Net significantly enhances predictions (refer to Fig. 6-(d,e,f), with lower and localized errors. Comparing the prediction of both velocity components, it shows that the adaptive model, i.e., the attention U-Net model, provides more accurate approximations for complex flow fields in LBM simulations, making it a promising alternative for improving prediction in complex flow fields.

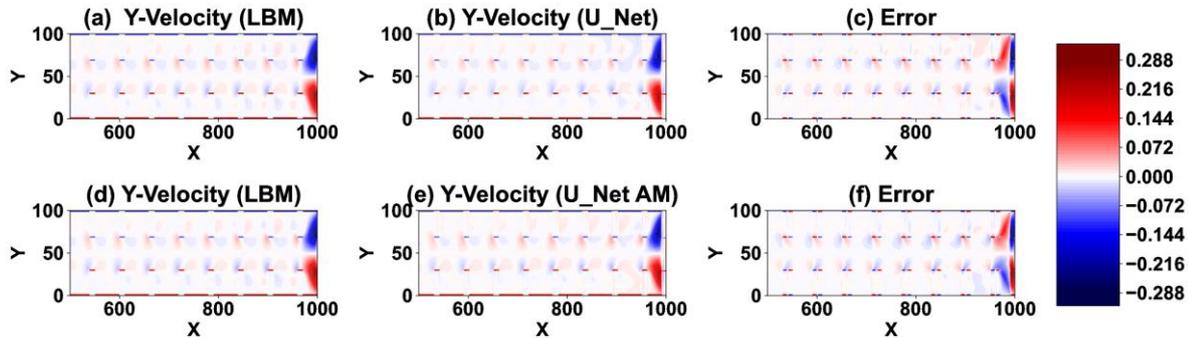

**Fig. 6** Prediction of y-velocity component using U-Net and attention U-Net (testing dataset).

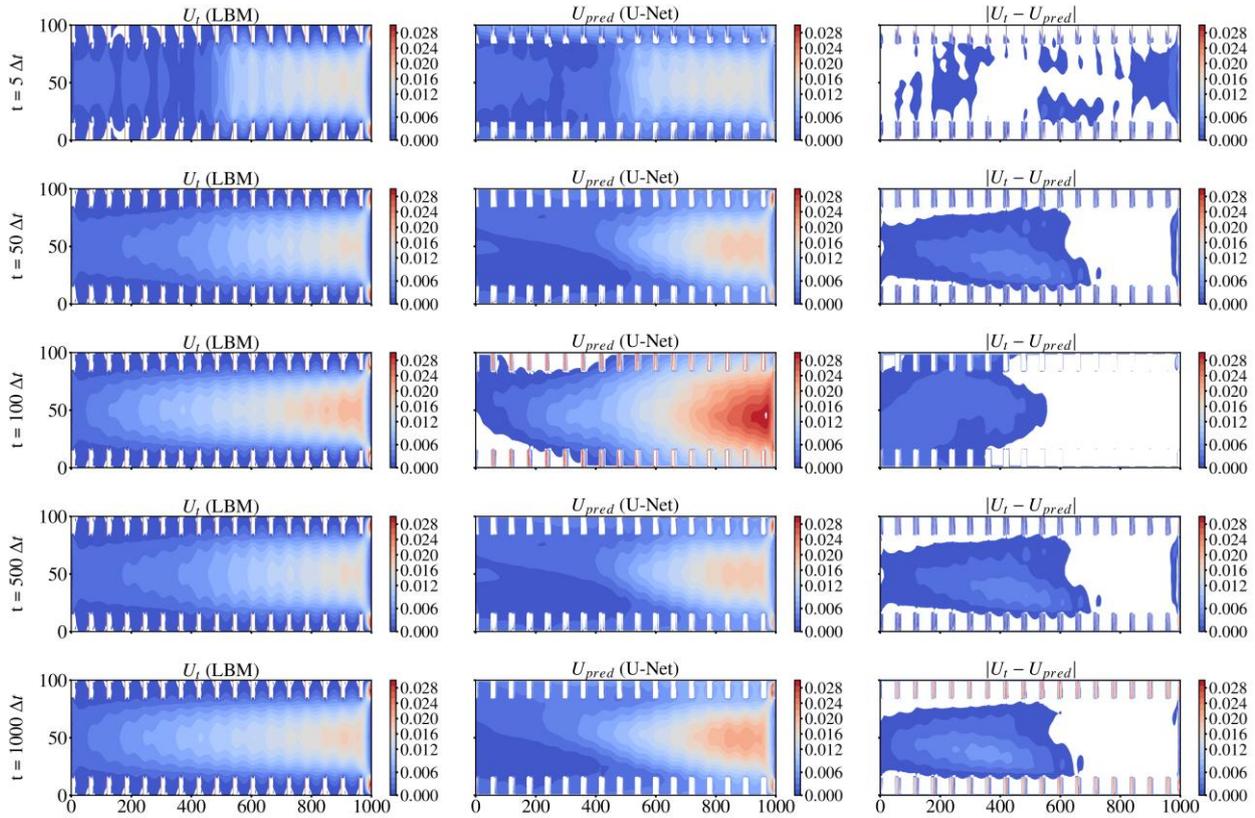

**Fig. 7** Prediction of unsteady velocity profiles using U-Net model for different timesteps a) Actual prediction

Figures 7 and 8 present the temporal analysis of a comprehensive fluid flow predictions using LBM and U- Net neural network across multiple time steps from $t = 5\Delta t$ to $1000\Delta t$. The visualization compares three key aspects at each time step: the actual LBM solutions ($U_t$), U-Net predictions ($U_{pred}$), and their absolute differences $|U_t - U_{pred}|$, with velocity magnitudes ranging from 0.000 to 0.028. At $t = 5\Delta t$, the flow exhibits initial development patterns with velocity peaks 0.024 near the obstacles. The intermediate time step $t = 50\Delta t$ shows established flow patterns with consistent velocity distributions of 0.012 to 0.016 in the channel center. The flow reaches a quasi-steady state at $t = 100\Delta t$, where the U-Net predictions demonstrate maximum deviations of 0.028 in the wake regions behind obstacles. The further time steps $t = 500\Delta t$ and $1000\Delta t$ maintain similar flow characteristics with velocity magnitudes between 0.006−0.020, while the absolute error distribution $| U_t - U_{pred} |$ reveals higher errors (up to 0.028) in the near-wall regions and texture interfaces. The spatial resolution of $100 \times 1000$ grid points effectively captures the flow features, including the periodic textures and associated wake patterns, with the error analysis highlighting the capability of the U-Net model to predict complex flow patterns while identifying regions of prediction uncertainty.

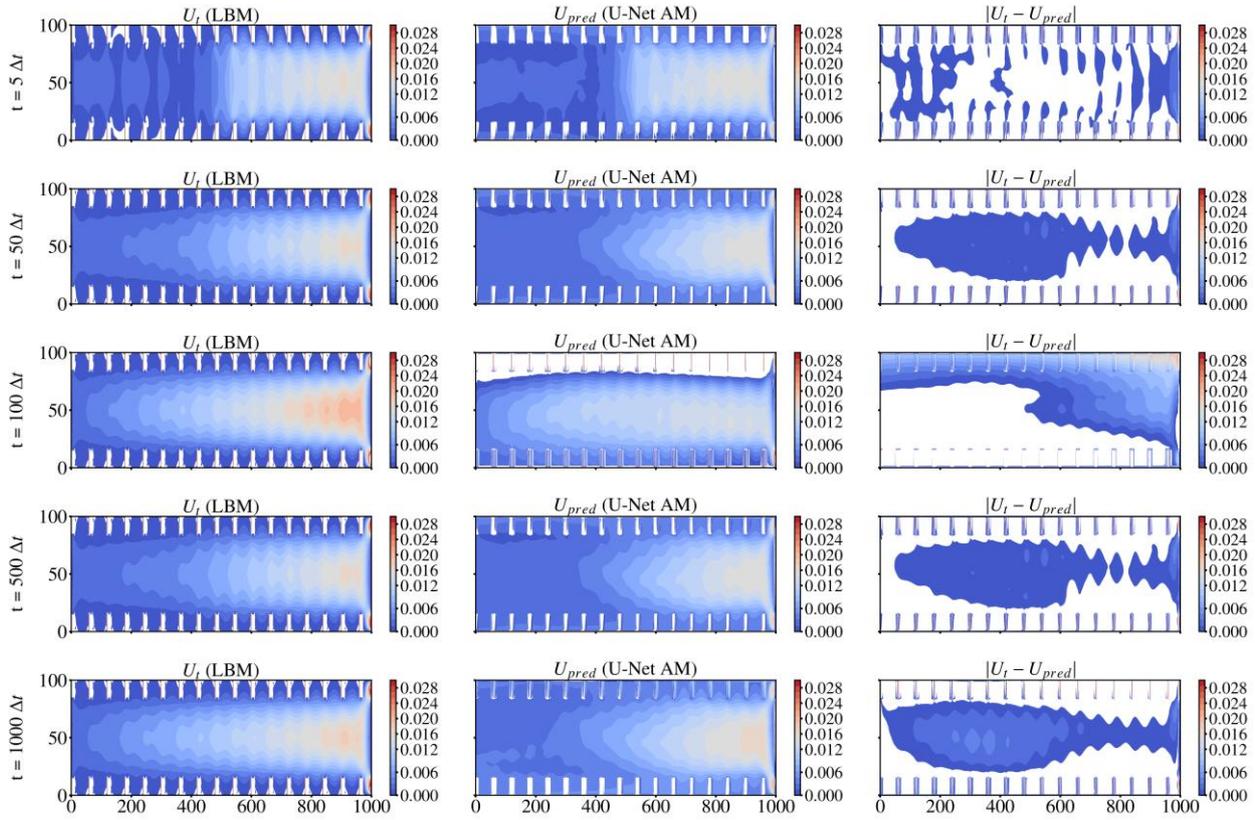

**Fig. 8** Prediction of unsteady velocity profiles using attention U-Net model for different timesteps
a) Actual prediction (LBM), and b) predicted (U-Net AM), and c) absolute errors.

Figure 9 presents a detailed comparison of fluid flow predictions between actual flow fields and U-Nets, showcasing both velocity components and vorticity fields across a domain. The actual streamwise velocity ranges from -0.055 to 0.44, while transverse velocity spans from -0.24 to 0.24, exhibiting periodic flow structures. The U-Net predictions show velocity ranges of -0.12 to 0.36 for the streamwise velocity component and -0.24 to 0.24 for a transverse component of velocity, while the attention U-Net demonstrates slightly improved predictions with ranges from -0.132 to 0.396 for $x$-velocity, and -0.22 to 0.275 for $y$-velocity. The vorticity fields $\omega = \dfrac{\partial V}{\partial x} - \dfrac{\partial U}{\partial y}$ reveal the significant errors with actual flow fields ranging from ±1.248, U-Net predictions ±1.025 and U-Net AM predictions ±1.2, indicating that the attention AM architecture regresses better vorticity patterns, particularly in capturing the periodic flow features and boundary interactions throughout the computational domain.

Figure 10-(a) displays a parabolic profile of fully developed flow at x/L = 0.3. The $x$-velocity attains a peak of roughly 0.2 at $Y = 50$, while the velocity diminishes to nearly zero at the boundaries at $Y = 20$ and $Y = 80$. The LBM approach demonstrates the most constant and smooth performance, closely followed by the Attention U-Net, which slightly underestimates the peak velocity. The U-Net exhibits a more pronounced deviation from the actual profile, particularly at the edges. The second and third numbers, which exhibit a maximum x-velocity of around 0.025, reveal more pronounced disparities between the approaches. The LBM consistently displays a smooth, parabolic trajectory, while both U-Net models show variability. Attention U-Net demonstrates enhancements over normal U-Net; nonetheless, both exhibit more volatility in the profile relative to LBM as shown in Fig. 10-(b). The coherence between LBM and attention U-Net is enhancing; nonetheless, the uncertainty of U-Net may restrict its reliability in more intricate velocity domains.

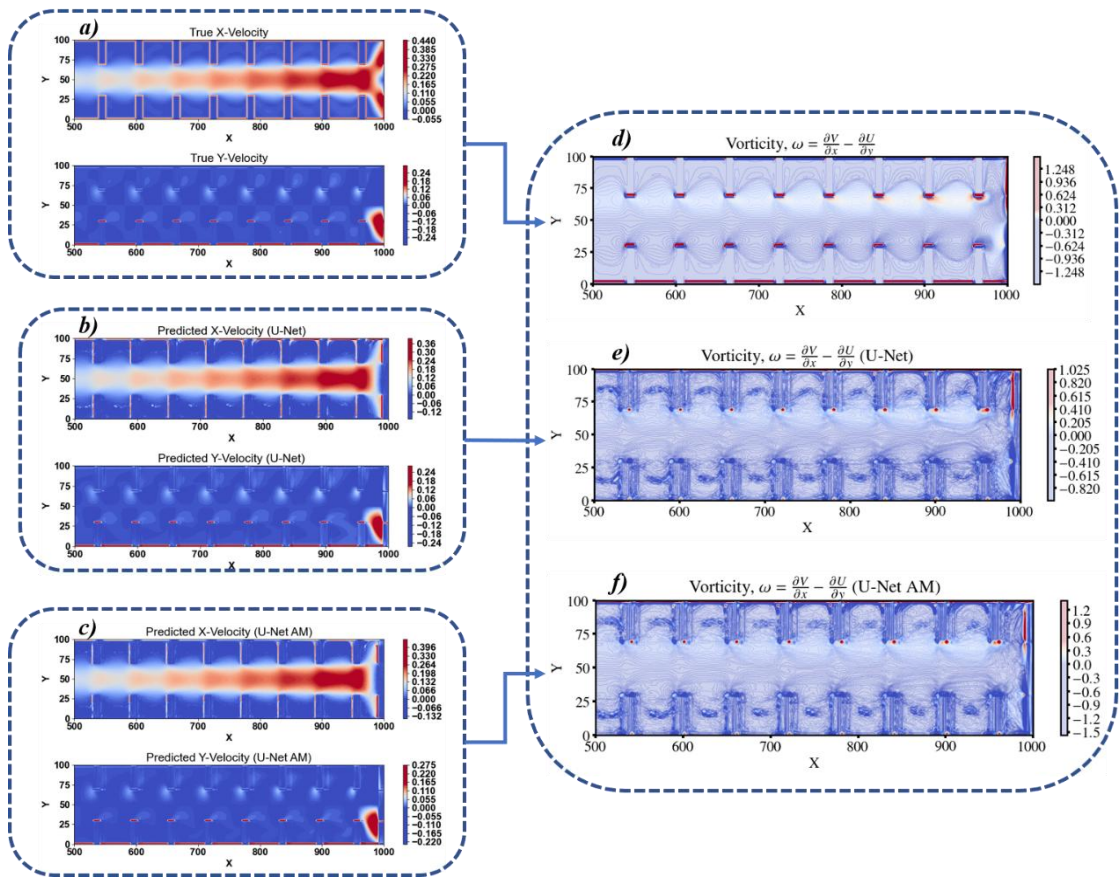

**Fig. 9** Vorticity profiles were generated using predicted velocity profiles, U-Net, and attention U-Net and comparing them with actual vorticity patterns generated from actual velocity components.

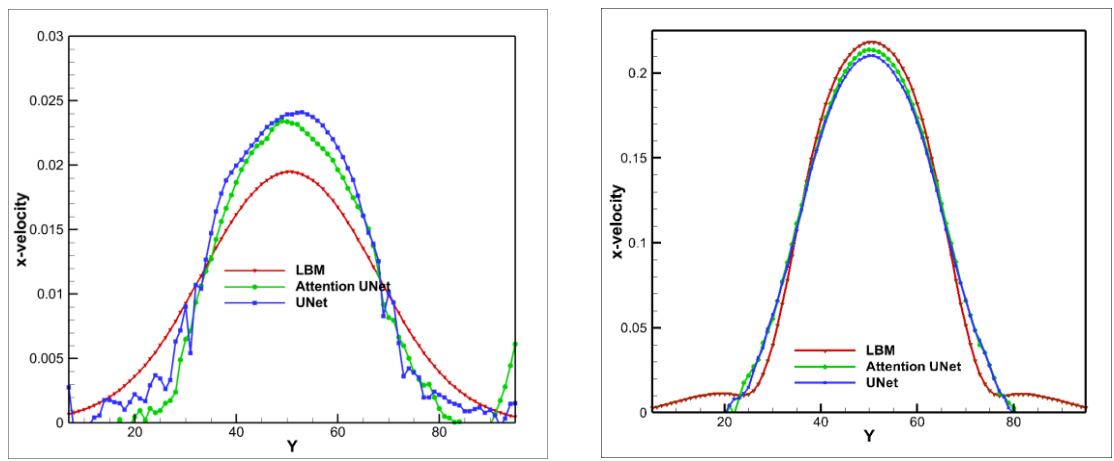

**Fig. 10** Prediction of velocity profiles using a) at x/L= 0.3, and b) at x/L= 0.8 for fluid flow datasets.

Figure 11 compares the performance of two models, U-Net and U-Net with attention mechanism, using four evaluative metrics; MAE, MSE, RMSE, and $R^2$. The results consistently demonstrate that U-Net AM outperforms the traditional U-Net model across all metrics. In terms of MAE, U-Net AM achieves a lower value, indicating more precise predictions on average. The MSE values for both models are similar, with U-Net AM showing a slight improvement, suggesting better performance in minimizing squared errors. RMSE follows a similar trend, with U-Net AM demonstrating a marginal enhancement over U-Net, signifying reduced average prediction errors. Most notably, the $R^2$ values for U-Net AM exceed 0.98, indicating excellent model performance, while the traditional U-Net model achieves performance up to 90%. These findings suggest that the integration of the attention mechanism in U-Net consistently improves predictive accuracy and generalization ability, although by a small margin, across all four measures. This improvement indicates that the attention mechanism effectively enhances the ability of the model to focus on relevant features, resulting in superior overall performance.

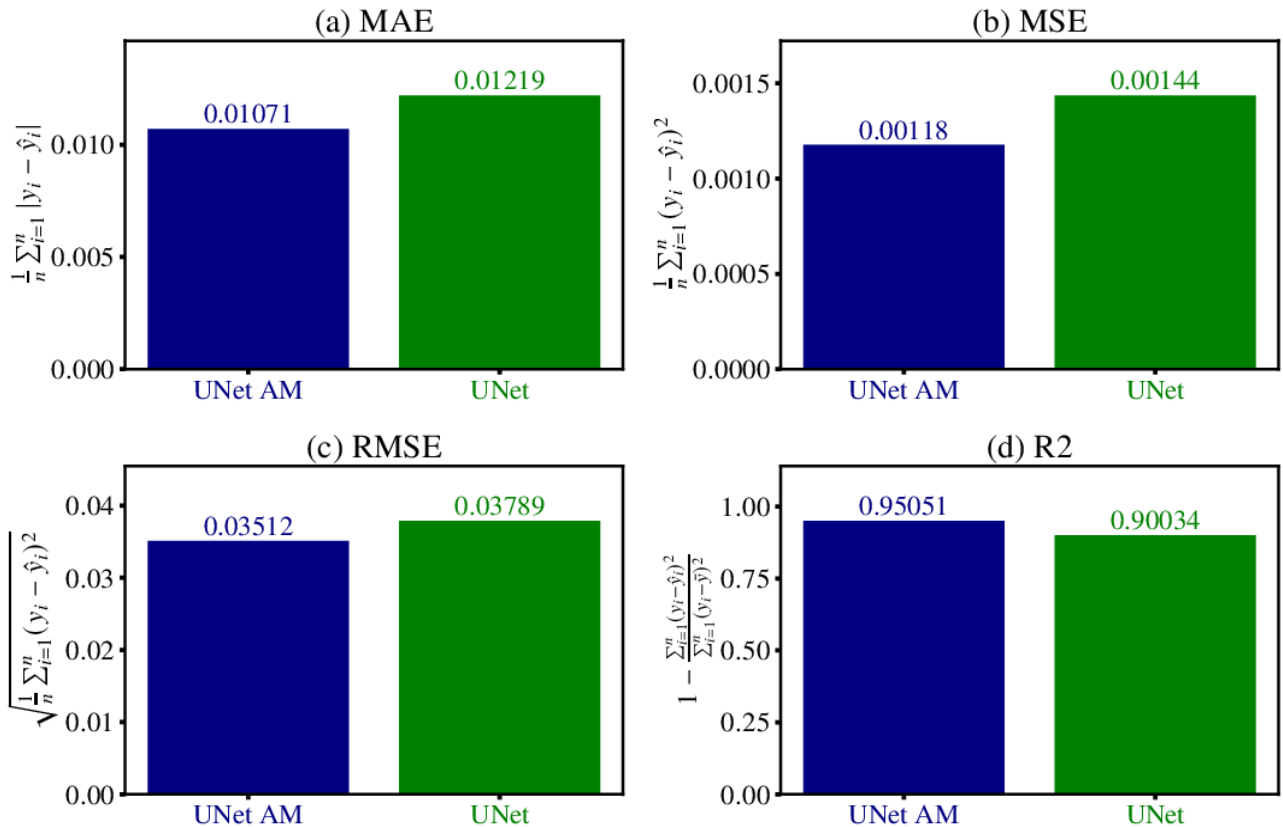

**Fig.11** Prediction evaluation using metrics for velocity profiles U-Net with attention and U-Net for fluid flow

## 6. CONCLUSIONS

In this study, we presented the systematic implementation of the U-Net deep learning architecture for a regression analysis to predict fluid flows through textured microchannels. Initially, the datasets were produced by the lattice Boltzmann method with the D2Q9 model. From the study, the U-Net with an attention mechanism demonstrates superior performance in predicting both velocity components and vorticity fields compared to the standard U-Net, with improved accuracy ranges and better preservation of flow features, particularly in capturing periodic patterns and boundary interactions across the computational domain. The provided U-Net with an attention mechanism can predict velocity magnitude and velocity components for the textured microchannels with an average error of 5.18%, and optimization is then able to reduce this error to 2.1%. The U-Net model with an attention mechanism (U-NetAM) consistently outperforms the standard U-Net model across all metrics, demonstrating improved accuracy and generalization. Although this study is focused on predictions of fluid flows through textured microchannels, these models give hope for the possibility of tackling more complex problems, such as the effect of hydrophobicity, flow through random rough surfaces at different Reynolds numbers, or 3D fields.

**Note:** The original paper was submitted to the ASTFE TFEC-2025 Conference which can be accessed through this link. http://dx.doi.org/10.1615/TFEC2025.ml.055517

## REFERENCES

[1] S. E. Game, M. Hodes, E. E. Keaveny, and D. T. Papageorgiou, "Physical mechanisms relevant to flow resistance in textured microchannels," Phys. Rev. Fluids, vol. 2, p. 094102, Sep 2017.

[2] S. Goli, S. K. Saha, and A. Agrawal, "Physics of fluid flow in an hourglass (converging-diverging) microchannel," Physics of Fluids, vol. 34, no. 5, 2022.

[3] Y. Q. Zu and Y. Y. Yan, "Single Droplet on Micro Square-Post Patterned Surfaces-Theoretical Model and Numerical Simulation," Scientific Reports, vol. 6, no. February 2015, pp. 1–12, 2016.

[4] M.A.Karnitz, M. C. Potter, and M. C. Smith, "An Experimental Investigation if Transition of a Plane Poiseuille Flu," tech. rep. 1974.

[5] R. M. Kiran and S. Chakraborty, "REVIEW PDMS microfluidics: A mini review," 2020.

[6] F. C. Yang and X. P. Chen, "Wetting failure condition on rough surfaces," Chinese Physics B, vol. 28, no. 4, 2019.

[7] D. Byun, J. Kim, H. S. Ko, and H. C. Park, "Direct measurement of slip flows in superhydrophobic microchannels with transverse grooves," Physics of Fluids, vol. 20, no. 11, 2008.

[8] C.Ishino, K. Okumura, and D. Qu´er´e, "Wetting transitions on rough surfaces," Europhysics Letters, vol. 68, no. 3, pp. 419–425, 2004.

[9] U. Akdag, M. A. Komur, and S. Akcay, "Prediction of heat transfer on a flat plate subjected to a transversely pulsating jet using artificial neural networks," Applied Thermal Engineering, vol. 100, pp. 412–420, 2016. Publisher: Elsevier Ltd.

[10] H. A. R. O L DSalwen and C. E. Grosch, "The stability of Poiseuille flow in a pipe of circular cross-section," tech. rep., 1972. Publication Title: J. Fluid Mech Volume: 54 Issue: 1.

[11] O.I. Vinogradova, "Slippage of water over hydrophobic surfaces," International Journal of Mineral Processing, vol. 56, pp. 31 60, Apr. 1999. Publisher: Elsevier Science Publishers B.V.

[12] S. L. Brunton, B. R. Noack, and P. Koumoutsakos, "Machine learning for fluid mechanics," Annual review of fluid mechanics, vol. 52, pp. 477–508, 2020. Publisher: Annual Reviews.

[13] S. Cai, Z. Wang, L. Lu, T. A. Zaki, and G. E. Karniadakis, "DeepM&Mnet: Inferring the electroconvection Multiphysics fields based on operator approximation by neural networks," Journal of Computational Physics, vol. 436, July 2021. arXiv: 2009.12935 Publisher: Academic Press Inc.

[14] X. Meng and G. E. Karniadakis, "A composite neural network that learns from multi-fidelity data: Application to function approximation and inverse PDE problems," Feb. 2019. arXiv: 1903.00104.

[15] F. Li, G. Ren, and J. Lee, "Multi-step wind speed prediction based on turbulence intensity and hybrid deep neural networks," Energy Conversion and Management, vol. 186, no. January, pp. 306–322, 2019. Publisher: Elsevier.

[16] S.Luo,J.Cui, M.Vellakal, J. Liu, E. Jiang, S. Koric, and V. Kindratenko, "Review and Examination of Input Feature Preparation Methods and Machine Learning Models for Turbulence Modeling," arXiv preprint arXiv:2001.05485, pp. 1–26, 2020. arXiv: 2001.05485.

[17] H. Xiao, J.-L. Wu, S. Laizet, and L. Duan, "Flows Over Periodic Hills of Parameterized Geometries: A Dataset for Data-Driven Turbulence Modeling From Direct Simulations," Computers & Fluids, p. 104431, 2020. arXiv: 1910.01264 Publisher: Elsevier Ltd.

[18] J. Kim and C. Lee, "Deep unsupervised learning of turbulence for inflow generation at various Reynolds numbers," Journal of Computational Physics, vol. 406, p. 109216, 2020. arXiv: 1908.10515 Publisher: Elsevier Inc.

[19] C. Guo, Y. Wang, Y. Han, M. Ji, and Y. Wu, "Unsteady flow-field forecasting leveraging a hybrid deep-learning architecture," Physics of Fluids, vol. 36, no. 6, 2024.

[20] L. Lu, X. Meng, Z. Mao, and G. E. Karniadakis, "DeepXDE: A deep learning library for solving differential equations," SIAM Review, vol. 63, no. 1, pp. 208–228, 2021. arXiv: 1907.04502 Publisher: Society for Industrial and Applied Mathematics Publications.

[21] S. Lee, F. Dietrich, G. E. Karniadakis, and I. G. Kevrekidis, "Linking Gaussian process regression with data-driven manifold embeddings for nonlinear data fusion," Interface Focus, vol. 9, no. 3, 2019. arXiv: 1812.06467 Publisher: Royal Society Publishing.

[22] H.Xie, D.Yang, N.Sun, Z.Chen, and Y.Zhang, "Automated pulmonary nodule detection in CT images using deep convolutional neural networks," Pattern Recognition, vol. 85, pp. 109–119, Jan. 2019. Publisher: Elsevier Ltd.

[23] P. P. Mehta, G. Pang, F. Song, and G. E. Karniadakis, "Discovering a universal variable-order fractional model for turbulent Couette flow using a physics-informed neural network," Fractional Calculus and Applied Analysis, vol. 22, pp. 1675–1688, Dec. 2019. Publisher: De Gruyter.

[24] M. Raissi, P. Perdikaris, and G. E. Karniadakis, "Physics-informed neural networks: A deep learning framework for solving forward and inverse problems involving nonlinear partial differential equations," Journal of Computational Physics, vol. 378, pp. 686–707, 2019. Publisher: Elsevier Inc.

[25] X. Jin, S. Cai, H. Li, and G. E. Karniadakis, "NSFnets (Navier-Stokes Flow nets): Physics-informed neural networks for the incompressible Navier-Stokes equations," Mar. 2020. arXiv: 2003.06496.

[26] C. W. Chang and N. T. Dinh, "Classification of machine learning frameworks for data-driven thermal fluid models," International Journal of Thermal Sciences, vol. 135, no. August 2018, pp. 559–579, 2019. arXiv: 1801.06621 Publisher: Elsevier.

[27] F. J. Mont´ans, F. Chinesta, R. Gomez-Bombarelli, and J. N. Kutz, "Data-driven modeling and learning in science and engineering," Comptes Rendus-Mecanique, vol. 347, no. 11, pp. 845–855, 2019. Publisher: Elsevier Masson SAS.

[28] T. Kr¨uger, H. Kusumaatmaja, A. Kuzmin, O. Shardt, G. Silva, and E. M. Viggen, The lattice Boltzmann method, vol. 10. Springer, 2017. Publication Title: Springer International Publishing Issue: 978-3.

[29] G.Meshram and S.Kondaraju, "Numerical Investigation of Wettability and its Effects on Flow through Textured Micro-channels using Lattice Boltzmann Method," in Proceedings of the 26thNational and 4th International ISHMT-ASTFE Heat and Mass Transfer Conference December 17-20, 2021, IIT Madras, Chennai-600036, Tamil Nadu, India, 2021.